# NbN/AlN/NbN Josephson junctions on sapphire for SIS receiver applications

M. Merker, C. Bohn, M. Völlinger, K. Ilin and M. Siegel

*Abstract*— The most developed integrated receivers for THz radiation nowadays are based on Josephson junction SIS devices. In this kind of devices, the highest receivable frequency is determined by the energy gap of the superconducting electrodes and limited to approximately 700 GHz in case of Nb/AlO$_x$/Nb multilayers.
We have developed a technology for NbN/AlN/NbN Josephson junctions on sapphire substrates which allow operation at frequencies above 1 THz. The trilayers are deposited in-situ in a 3-chamber DC-magnetron sputtering system at temperatures as high as 775 °C. Each layer is reactively sputtered in an argon and nitrogen atmosphere. By variation of the partial gas pressures and discharge current of the plasma, optimum deposition conditions for the tri-layers were found. So far, gap voltages as high as 5.1 mV at 4.2 K have been achieved which allows the operation frequency of the JJ devices to exceed 1 THz.

*Index Terms*—Josephson junctions, superconducting devices, superconducting films, submillimeter wave integrated circuits

## I. Introduction

INTEGRATED receivers based on Superconductor-Insulator-Superconductor (SIS) structures offer the possibility of very sensitive detection of radiation in the sub-mm wavelength range. In the last decades, this kind of receivers has advanced to levels of quality and integration which allows the detection of frequencies as high as 700 GHz with sensitivities close to the quantum limit [1-6]. Most of the realized devices were based on the well-established Nb/Al-AlO$_x$/Nb technology. The electrodes made from niobium exhibit a gap voltage of 2.9 mV, resulting in an upper frequency limit of 700 GHz. In order to overcome this limit, niobium nitride (NbN) or niobium titanium nitride was used in some cases as material of choice for one of the electrodes [7-12]. However, in order to achieve ultimate performance, both of the electrodes need to be of materials with high gap voltages, e.g. NbN. Since the lattice constant of NbN ($a_{NbN}$ = 0.439 nm) almost perfectly fits the one of Magnesium oxide (MgO) ($a_{MgO}$ = 0.42 nm), NbN/AlN/NbN Josephson junctions (JJs) of highest quality are realized on MgO [12, 13]. Achieving gap voltages higher than 4.5 mV on substrates other than MgO is challenging because of the much larger lattice mismatch (e.g. $a_{Sapphire}$ = 0.476 nm) [14-16]. However, other substrates like sapphire are preferable over the highly hygroscopic MgO, since the absorbed water decreases the overall transmission of THz radiation through the MgO substrate. There are two major properties, the trilayer structures for Josephson devices have to fulfill, in order to achieve the required quality: the energy gaps need to be high enough for the operation frequency and the thin barriers need to uniformly isolate the electrodes from each other.

The energy gaps of the superconductors determine the maximum operation frequency, i.e. $2\Delta = h \cdot \nu_{max}$, where $\Delta$ is the energy gap of the superconductor, $h$ is the Planck constant and $\nu_{max}$ is the maximum operation frequency of the device. For a target frequency of 1 THz, an energy gap of at least 4.2 meV is required. For very low temperatures compared to the critical temperature of the superconductor, the energy gap can be calculated from $\Delta(T=0) = S \cdot k_B \cdot T_c$, where $k_B$ is the Boltzmann constant, $T_c$ is the critical temperature, and $S$ is a material dependent parameter. According to BCS theory, this parameter has a constant value of 1.764 in the case of weakly coupled superconductors. NbN however is a superconductor with strong coupling. Therefore, this parameter can differ from this value. In [17], values as high as 2.2 have been reported for NbN films. Assuming an almost constant material parameter $S$ for the material system, maximizing the superconducting energy gap is equivalent to the optimization of the critical temperature. Investigations on the fabrication of NbN films showed, that the critical temperature is strongly influenced by the deposition temperature [18], the deposition current [19], and the relative partial pressures of argon and nitrogen [20,21], i.e. parameters that influence on the stoichiometry and crystalline structure of the films.

The aluminum nitride (AlN) barrier in between the electrodes is only few nm thick but still has to fully isolate the electrodes from each other uniformly. Therefore, the surface of the lower electrode has to be smooth on a sub-nm scale. The barrier thickness is inversely proportional to the critical-current density ($j_c$) of the trilayer. For SIS mixers, high $j_c$ values are favorable since both, noise temperature and IF bandwidth improve. Therefore, $j_c$ needs to be as high as possible.

In order to obtain high quality JJs, a careful investigation on the deposition technology is required. Here, we present the optimization process of NbN/AlN/NbN Josephson junctions on sapphire substrates with the main focus on high superconducting energy gaps enabling the operation frequency to exceed 1 THz.

The authors are with the Institute of micro- and nano-electronic systems (IMS) at Karlsruhe Institute of Technology (KIT), 76187 Karlsruhe, Germany
(e-mail: michael.merker@kit.edu).

## II. Josephson junctions on sapphire

The used deposition system is a self-developed 3-chamber reactive magnetron sputtering system. One chamber is the load lock with RF pre-cleaning, the other chambers contain a niobium and an aluminum target, respectively. During deposition of the NbN/AlN/NbN trilayer, the R-cut sapphire substrates were placed on a heater kept at a temperature of 775 °C. The movable heater stage allows a deposition of the complete multi-layer structure onto heated substrates without breaking the heating process. Due to the reactive nature of the deposition process, the deposition parameters cannot be changed separately, but are dependent on each other. In order to obtain the influence of a single parameter, another parameter needs to be scaled.

### A. Optimization of the electrodes

A major property of the superconducting electrodes is their energy gap and thus their critical temperature. Fig. 1 shows the dependence of $T_c$ on the nitrogen partial pressure $p_{N2}$. The partial pressure of the nitrogen is swept, while the argon partial pressure $p_{Ar}$ is kept constant at $4.5 \cdot 10^{-3}$ mbar. The deposition current is scaled as follows. In the plasma *IV*-characteristic, there is a region with a negative differential resistance (similar to fig. 1 in [19]). The optimum deposition current is expected to be close to or within this region. Changing the nitrogen partial pressure also changes the *IV*-characteristic, and the deposition current needs to be adopted so as to stay in the same point relative to this region of negative differential resistance. The critical temperature reaches its maximum at the nitrogen partial pressure of $1.5 \cdot 10^{-3}$ mbar, i.e. 25 % nitrogen content of the whole gas mixture. The stoichiometry of the NbN films is furthermore influenced by the deposition current. Fig. 2 shows the dependence of the critical temperature on deposition current. The gas mixture is fixed at the optimum combination of partial pressures of argon ($p_{Ar} = 4.5 \cdot 10^{-3}$ mbar) and nitrogen ($p_{N2} = 1.5 \cdot 10^{-3}$ mbar), which have been found from the investigation described above.

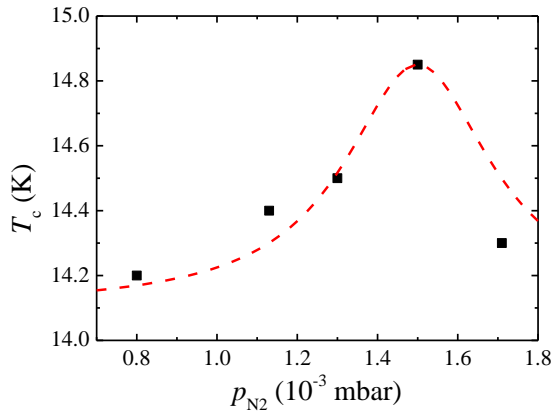

Figure 1: The dependence of the critical temperature on the partial pressure of nitrogen, maintaining a constant argon pressure of $p_{Ar} = 4.5 \cdot 10^{-3}$ mbar. The dashed line is to guide the eye. For each nitrogen partial pressure, the current has been adjusted.

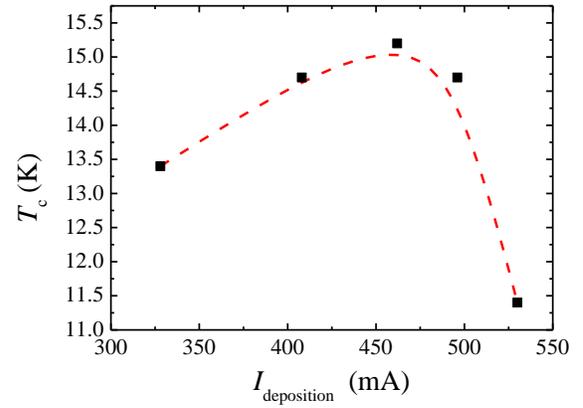

Figure 2: Dependence of the critical temperature on the deposition current with $p_{Ar} = 4.5 \cdot 10^{-3}$ mbar, $p_{N2} = 0.45 \cdot 10^{-3}$ mbar. The dashed line is to guide the eye. The maximum critical temperature is 15.2 K.

The maximum critical temperature of 15.2 K was achieved at $I_{deposition} = 462$ mA. According to our estimations, this value of $T_C$ corresponds to an energy gap of 5.7 meV.

The second major property of the electrodes, particularly of the lower one, is the surface roughness which is mostly determined by the total pressure during deposition [18]. In Fig. 3, scanning electron microscopy images of the surface of NbN films are shown. These films are deposited at different argon pressures of $4.5 \cdot 10^{-3}$ mbar (Fig. 3a) and $1.6 \cdot 10^{-3}$ mbar (Fig. 3b). In order to keep the critical temperature constant, the other parameters have to be changed accordingly: the nitrogen content is kept at 25 % of the total pressure, i.e. $p_{N2} = 1.5 \cdot 10^{-3}$ mbar and $0.54 \cdot 10^{-3}$ mbar. The current was again scaled to stay on the same relative point in the *IV*-characteristic of the plasma, i.e. $I_{deposition} = 462$ mA and 192 mA. Crystallites with sharp edges are clearly visible in Fig. 3a, indicating a strong polycrystalline growth of the film. By reducing the total pressure, the occurrence of the crystallites decreases, indicating a smoother surface, until they vanish completely. The film shown in Fig. 3b, demonstrates an almost perfectly smooth surface without any sign of formation of the crystallites. Therefore the total pressure needs to be as low as possible in order to achieve surfaces of the NbN layers that are suitable for subsequent growth of nm-thick AlN isolation barrier. The lowest argon pressure in our system at which the plasma is stable is $1.4 \cdot 10^{-3}$ mbar.

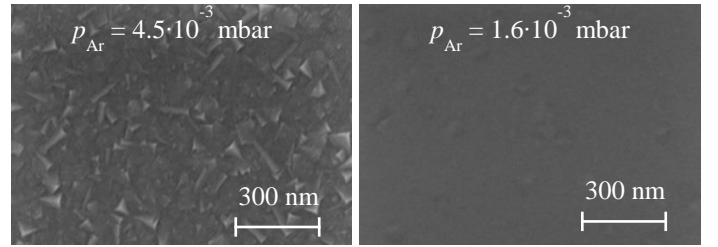

Figure 3: Scanning electron micrographs showing the surface of NbN films deposited at different total pressures.
a) $p_{Ar} = 4.5 \cdot 10^{-3}$ mbar, $p_{N2} = 1.5 \cdot 10^{-3}$ mbar, $I_{deposition} = 462$ mA
b) $p_{Ar} = 1.6 \cdot 10^{-3}$ mbar, $p_{N2} = 0.54 \cdot 10^{-3}$ mbar, $I_{deposition} = 192$ mA

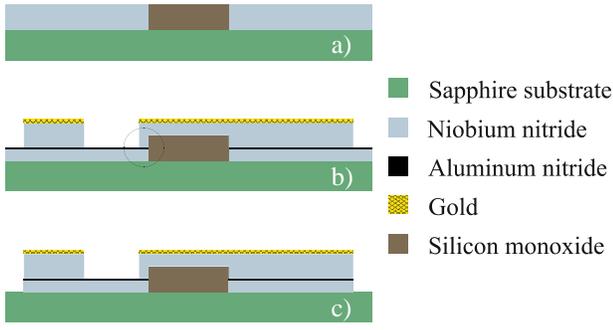

Figure 4: Schematics showing the cross sections at intermediate fabrication steps. The black circle in (b) marks the junction area.

The scaled nitrogen pressure is thus $0.47 \cdot 10^{-3}$ mbar and the optimum deposition current for this pressure is 160 mA.

*B. JJ fabrication and optimization of the barrier*

The barrier optimization was done by investigation of the DC-properties of Josephson junctions fabricated in the following sequence. The thickness of each NbN layer is 200 nm and approximately 2 nm for the AlN layer. Since the barrier as well as the electrodes is deposited in a reactive sputtering process, a certain gas settling time $t_\text{gas-setup}$ is required to set up the atmosphere inside the process chamber. When the pressure is stable, the plasma is ignited and stabilized. Then, the deposition is started by opening the shutter. The NbN layers are deposited according to the optimized parameters discussed in the previous section. The barrier deposition parameters were adopted from previous work [18] i.e. $p_\text{Ar} = 3 \cdot 10^{-3}$ mbar, $p_\text{N2} = 1 \cdot 10^{-3}$ mbar and $I_\text{deposition} = 200$ mA. The trilayers were then patterned using a refined version of our standard technological process for JJ test structures, described in detail in [18]. This process allows for short processing times, maintaining a high quality of Josephson junctions and will be explained in the following. The 10 x 10 mm$^2$ sapphire substrates are fully covered with the trilayer. Trenches of 20 μm in width and 9050 μm in length are defined by photolithography and etched down to the substrate using reactive-ion etching (RIE) and ion-beam etching (IBE) techniques. The sidewalls of the trenches are then oxidized and the trenches are filled with a thermally evaporated Silicon monoxide to the trilayer height. After lift-off of the SiO outside of the trenches, a flat surface of the chip is obtained (Fig. 4a). A subsequent lithography defines the junction dimensions after which the 300 nm thick NbN wiring layer is deposited, with a 20 nm thick gold film on top and lifted off. The junctions are then structured using RIE (Fig. 4b). The top layer of Au hereby serves as a hard mask. Separating all devices on the chip from each other is done with another lithography and RIE (Fig. 4c). After stripping of the resist, the junctions can be measured. All JJ measurements are performed in a liquid-helium transport dewar at a temperature of 4.2 K. Fig. 5 shows the comparison of *IV*-characteristics of two junctions with identical electrodes but different barriers: the deposition time $t_\text{dep,AlN}$ was varied slightly, $t_\text{gas-setup}$ was changed dramatically.

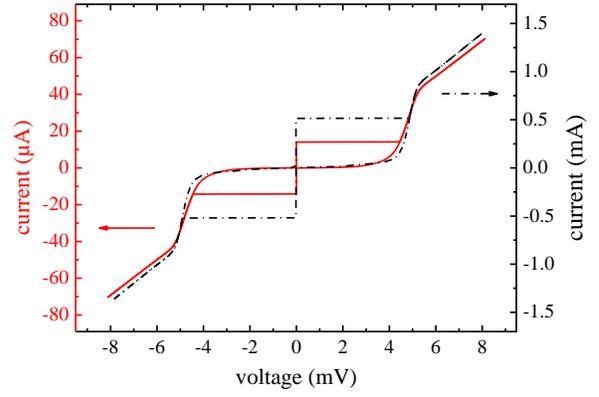

Figure 5: Current-voltage characteristics of JJs with different barriers and identical design area of 36 μm$^2$. Both, the gas settling time as well as the AlN deposition time are different. The quality parameters of the JJs clearly improve for shorter settling times (see table I). The solid red curve represents HN06, the dashed black curve represents HN26.

TABLE I
JUNCTION QUALITY PARAMETERS

| Name | $t_\text{gas-setup}$ | $t_\text{dep,AlN}$ (sec) | $I_\text{dep.}$ (mA) | $j_c$ (A/cm$^2$) | $V_\text{gap}$ (mV) | Subgap ratio | $I_C R_N$ (mV) |
|---|---|---|---|---|---|---|---|
| HN06 | 5 min | 25 | 200 | 39 | 4.6 | 18 | 1.5 |
| HN26 | 15 sec | 35 | 200 | 1435 | 4.9 | 13 | 2.7 |

Table I: Comparison of two trilayers with different gas settling times and different barrier deposition times. The given values are color coded according to the *IV*-curves in Fig. 5.

The particular values for the mentioned times along with the measured quality parameters are given in Table I.

The critical-current density $j_C$ is the ratio of the critical current and the design area of the JJ. The gap voltage $V_g$ is the voltage at 50 % of the tunnel-current jump. The subgap ratio $R_\text{sg}/R_N$ is the ratio between the subgap resistance $R_\text{sg}$ and the normal state resistance $R_N$. The subgap resistance hereby is taken at a voltage of 70 % of the gap voltage. Finally $V_m$ is the product of the critical-current and the subgap resistance. If we assume identical electrodes and ideal interfaces, the critical current density of the trilayer should only be defined by the barrier thickness, which is set by deposition time. Since $t_\text{dep,AlN}$ was longer for HN26, a slightly lower $j_c$ was expected. However, the measurements show a significant increase in $j_c$, indicating that the effective barrier thickness does not correlate with $t_\text{dep,AlN}$. We assume that the nitrogen in the chamber during $t_\text{gas-setup}$ modifies the surface of the bottom electrode, resulting in a larger effective barrier thickness. In order to prove this assumption, further investigation is required. For all other trilayers shown here, $t_\text{gas-setup}$ was closely monitored and kept as short as possible.

### III. JJ MEASUREMENTS AND DISCUSSION

The optimum deposition parameters for each layer were found in the optimizations described in the previous section. The electrodes are deposited at $p_\text{Ar} = 1.4 \cdot 10^{-3}$ mbar, $p_\text{N2} = 0.47 \cdot 10^{-3}$ mbar and $I_\text{deposition} = 160$ mA. The barrier is deposited with $p_\text{Ar} = 3 \cdot 10^{-3}$ mbar, $p_\text{N2} = 1 \cdot 10^{-3}$ mbar at $I_\text{deposition} = 200$ mA. With this full set of deposition parameters, various trilayers were fabricated and measured.

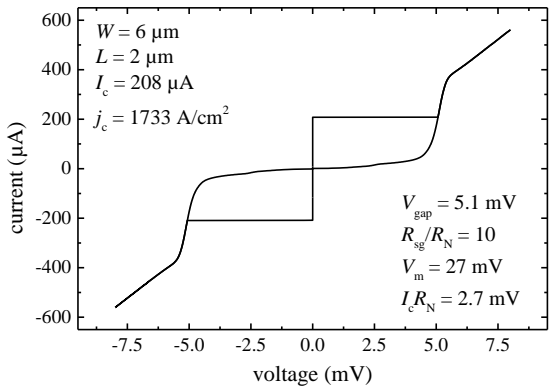

Figure 6: IV characteristic of a JJ fabricated with the optimized parameters. This JJ showed the highest gap voltage of 5.1 mV.

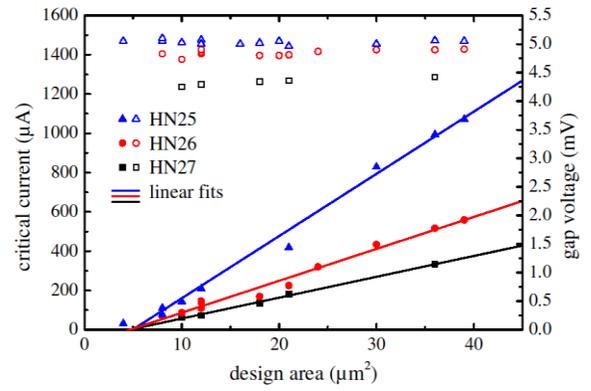

Figure 8: Dependence of the critical current (closed symbols) on the design area of the JJ. The critical current scales linearly with the design area. The open symbols show the gap voltage of the different JJs.

Due to small variations in deposition conditions of the electrodes, the critical temperature changes slightly for different trilayers. This results in a variation of the gap voltage from trilayer to trilayer between 4.3 mV and the maximum value of 5.1 mV (see Fig. 6). The gap voltage is thus sufficient for operation of the devices beyond 1 THz. However, in order to achieve high quality receiver devices, some of the other quality parameters (inset of the graph) need further improvement. The achieved values of $I_cR_N$ ranging from 2.7 mV to 3.9 mV are acceptable. The highest $j_c$ measured so far is 3.7 kA/cm$^2$. This parameter was not the primary goal of this study and needs further improvement. The subgap ratio is acceptable with values ranging from 9 to 15. However it should be increased to values exceeding 20 in order to obtain good noise properties of receiver devices. The homogeneity of the current distribution within the junction was investigated by the modulation of the critical current due to an externally applied magnetic field. The result is depicted in Fig. 7 along with the Fraunhofer fit. The critical current almost perfectly follows the Fraunhofer distribution, proving that the current density is homogenous across the junction. The investigation of the homogeneity of the trilayer across the chip is done by the measurement of several junctions located in different positions on the chip. Fig. 8 shows the critical currents of different junctions vs their area. With minor deviations, the critical currents follow a linear fit.

The offset on the x-axis results from the fact that the areas of the JJs were not measured, but the design areas were used. Due to imperfection of photolithography and the isotropic nature of the reactive ion etching process, the junction area is likely to be smaller than expected, resulting in this offset. Along with the critical currents, the measured gap voltages are shown in Fig. 8. As mentioned above, the gap voltages vary for different trilayers. However, on one chip, the gap voltage stays constant except for small fluctuations. These results indicate a homogeneous trilayer deposition across the whole chip area (10 x 10 mm$^2$).

IV. CONCLUSION

We have optimized the deposition process for NbN/AlN/NbN trilayer on sapphire substrates. Critical temperatures as high as 15.2 K has been reached by optimization of the stoichiometry and crystalline structure of the NbN films of bottom and top electrodes. By reducing the total pressure to the minimum stable point, the surface roughness of the films was decreased to a level which allows the fabrication of Josephson junctions with aluminum nitride barriers. Gap voltages as high as 5.1 mV were achieved. The reduction of the gas settling time for barrier deposition resulted in a sharp current rise and acceptable subgap ratios at critical current densities exceeding 1 kA/cm$^2$. The homogeneity of the current density distribution was proven by $I_c(H)$ measurements showing an excellent match with the Fraunhofer pattern. The homogeneity of the trilayers across the chip is shown by the distribution of the critical currents and uniformity of the gap voltage across the chip. For the realization of high-quality Josephson junctions for SIS receiver applications, the subgap ratio and critical current density should be improved. The achieved gap voltages are sufficient to create SIS devices suitable for operation frequencies exceeding 1 THz.

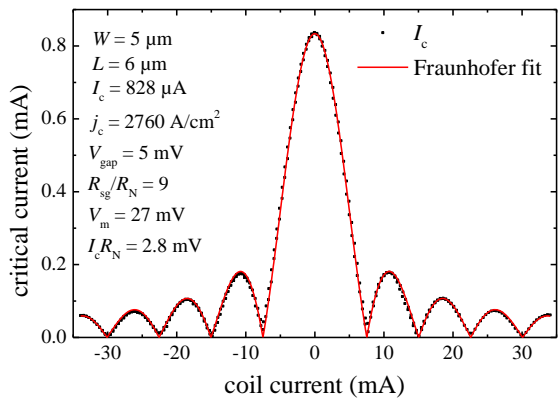

Figure 7: Dependence of the critical current on an externally applied magnetic field. The measurement points almost perfectly follow the Fraunhofer fit (red curve), proving the homogeneity of the current density through the JJ.

ACKNOWLEDGMENT

The authors would like to thank K.H. Gutbrod and A. Stassen for their technical and F. Ruhnau for his IT support.